\documentstyle[prl,aps,psfig,epsf]{revtex}

\begin{document}


\title{On a Connection Between Nonlinear Response
to an External Field and Equilibrium Properties of Systems with
Interaction: Comment on cond-mat/9711226 by Kopietz and V\"olker}

\author{V.E.Kravtsov  \\
International Centre for Theoretical Physics
\\ P.O.Box 586, 34100 Trieste, Italy\\
and Landau Institute of Theoretical Physics,\\
2 Kosygina str., 117940 Moscow, Russia.\\
and\\
V.I. Yudson\\
Institute of Spectroscopy,\\
Russian Academy of Sciences,\\
142092 Troitsk, Moscow r-n, Russia \\
and CCP, Physics \& Astronomy Building,\\
The University of
Western Ontario,\\ London, Ontario, Canada N6A 3K7.}

\maketitle
\begin{abstract}
We demonstrate the
danger of a naive identification of kinetic
coefficients in the free-electron response expansion
in powers of an {\it external} electric field and
coefficients in the expansion of the corresponding equilibrium
quantity in powers of the fluctuating {\it intrinsic} electric field.
A particular example of the direct current generation
by an AC electric field in a mesoscopic system of non-interacting
electrons vs. the Fock contribution to the equilibrium persistent
current in a system of interacting electrons is considered.
We comment on a recent paper by Kopietz and V\"olker,
cond-mat/9711226.
\end{abstract}

The problem of an adequate theoretical description of electron-electron
interaction and electron relaxation processes in disordered mesoscopic
conductors has become a very hot topic nowadays \cite{GEN}.
The complexity of the problem stimulates the search for approaches
which incorporate general relationships between kinetic and
equilibrium properties of the system.
One may believe that the knowledge of the nonlinear response to an
external field supplemented by the field correlation functions obtained,
e.g., with the use of the fluctuation-dissipation theorem (FTD)
\cite{LL} is helpful for solving the problem.

The purpose of this Letter is to demonstrate how dangerous might be
a naive way of applying FDT-like relations for description of
systems of {\it interacting} electrons, when the coefficients of expansion
of
an equilibrium quantity in powers of the fluctuating intrinsic
electric field are identified with the kinetic coefficients in the
free-electron response expansion in powers of an external
electric field.

The reason is a deep physical difference between the dissipative
nature of the causal response to an external field and
the ``non-causal'' nature of all the interaction processes in the
equilibrium.

As an actual example we consider the nonlinear generation of the
direct current $I^{(2)}_{DC}$ of {\it non-interacting} electrons
in a mesoscopic ring of the circumference $L$ pierced by a
constant magnetic flux $\phi$ and an AC
flux $\phi_{\omega}$ that produces an electric field ${\cal E}(\omega)=
(i\omega/Lc)\phi_{\omega}$. This kinetic problem has been studied
earlier
in Ref.\cite{KY},\cite{AK} and an expression for the  nonlinear
conductance $\sigma^{(2)}_{kin}(\omega, -\omega)$
defined by
\begin{eqnarray}\label{I2}
I^{(2)}_{DC} =
\sigma^{(2)}_{kin}(\omega, -\omega)|{\cal
E}_{\omega}|^2,
\end{eqnarray}
has been obtained. A peculiar feature of the result of
Refs.\cite{KY},\cite{AK} is that the nonlinear conductance
$\sigma^{(2)}_{kin}(\omega, -\omega)$
does not vanish exponentially in the limit of high frequencies $\omega\gg
E_{c}$, where $E_{c}=D/L^2$ is the Thouless energy and $D$ is the
diffusion coefficient. This is in a striking contrast to a similar
mesoscopic
effect of the Aharonov-Bohm oscillations of the linear conductance
\cite{AB}.

Recently there was an attempt by Kopietz and V\"olker \cite{KV} to
relate the nonlinear
conductance $\sigma^{(2)}_{kin}(\omega, -\omega)$ with the
coefficient
$\sigma^{(2)}_{eq}({\bf q},i\omega_{n})$ in the expression for the Fock
(exchange) contribution to the equilibrium persistent current
in the system of {\it interacting} electrons:
\begin{equation}\label{IF}
I^{(F)}_{PC} =- \frac{T}{2}\sum_{{\bf q}, n}
\sigma^{(2)}_{eq}({\bf q}, i\omega_n)\frac{\langle({\bf q \cdot
E_q}(i\omega_n))
({\bf q\cdot E_{-q}}(-i\omega_n))\rangle}{|{\bf q}|^2}.
\end{equation}
Here $\omega_{n}$ are bosonic Matsubara frequencies $\omega_{n}=2\pi Tn$,
and $\langle E^{\alpha}_{{\bf
q}}(i\omega_{n})\,E^{\beta}_{{-\bf
q}}(-i\omega_{n})\rangle=q_{\alpha}q_{\beta}\,D({\bf q},i\omega_{n})
$ is
proportional to a disorder-averaged Green's function of longitudinal
photons \cite{AGD} $D({\bf q},i\omega_{n})=4\pi /q^2
\varepsilon({\bf
q},i|\omega_{n}|)$ in a media with the dielectric function
$\varepsilon({\bf
q},\omega)$.

One can transform Eq.(\ref{IF}) to an integral over real frequencies
by representing
the sum in terms of a contour integral.
\begin{equation}
\label{acon1}
I^{(F)}_{PC}=I_{PC}^{st} + \sum_{{\bf q}}
\int_{0}^{+\infty}\frac{d\omega}{2\pi
}\,I^{(F)}_{PC}({\bf q},\omega),
\end{equation}
where  $I_{PC}^{st}$ is a ``static'' contribution given by
the $\omega_{n}=0$ term of Eq.(\ref{IF}) and
\begin{equation}
\label{acon}
I^{(F)}_{PC}({\bf q},\omega)=\coth\left(
\frac{\omega}{2T}\right)\,\Re\left[i\sigma_{eq}^{(2)}({\bf
q},\omega)\,q^2 D({\bf q},\omega)
\right].
\end{equation}
Here
$\sigma_{eq}^{(2)}({\bf
q},\omega)=[\sigma_{eq}^{(2)}({\bf
-q},-\omega)]^{*}$ and $D({\bf q},\omega)=[D({\bf
-q},-\omega)]^{*}
$ are analytical continuations of $\sigma^{(2)}_{eq}({\bf q}, i\omega_n)$
and $D({\bf q},i\omega_{n})$ in the upper half-plane of $\omega$.

Eq.(\ref{acon}) can be further simplified in the  Ohmic
regime characterized by a constant conductivity $\sigma_{0}$, where
$\varepsilon({\bf q},\omega)=4\pi\sigma_{0}/(-i\omega)$. In this case
$D({\bf q},\omega)$ is purely imaginary and is related with the
spectral density of the fluctuating intrinsic electric filed
$\langle|{\bf E}_{{\bf q},\omega}|^2 \rangle$ by the FDT \cite{LL}:
\begin{equation}
\label{FDT}
\langle |{\bf E}_{{\bf q},\omega}|^2\rangle=
i\coth\left(\frac{\omega}{2T}\right)\,q^2 D({\bf q},\omega) =
\coth\left(\frac{\omega}{2T}\right)\,\frac{\omega}{\sigma_{0}}.
\end{equation}
Thus we arrive at an expression analogous to Eq.(\ref{I2}):
\begin{equation}
\label{rf}
I^{(F)}_{PC}({\bf q},\omega)=\Re\, \sigma_{eq}^{(2)}({\bf
q},\omega)\,\langle|{\bf E}_{{\bf q},\omega}|^2 \rangle.
\end{equation}
It is tempting to interpret Eq.(\ref{rf}) as a ``nonlinear response''
to the fluctuating intrinsic field ${\bf E}_{{\bf q},\omega}$
in the system of interacting electrons, by an analogy with the response
Eq.(\ref{I2}) to an external field ${\cal E}_{{\bf q},\omega}$
applied to a free electron system.

Then one may believe  \cite{KV} that
there is a simple relationship between $\sigma^{(2)}_{kin}(\omega,-\omega)$
in Eq.(\ref{I2}) and
$\sigma_{eq}^{(2)}({\bf
q},\omega)$ in Eq.(\ref{rf}):
\begin{eqnarray}\label{SKV}
\sigma^{(2)}_{kin}(\omega,-\omega) = \mbox{\rm lim}_{{\bf
q}\rightarrow
0}\,\,
\Re \sigma^{(2)}_{eq}({\bf q}; \omega).
\end{eqnarray}

However, this leads to a dramatic discrepancy with
Ref.\cite{KY,AK}, since $\sigma^{(2)}_{eq}({\bf q}; \omega)$
appears to be \cite{KV} exponentially small 
at $\omega\gg
E_{c}$, while the $\sigma^{(2)}_{kin}(\omega,-\omega)$ decreases only as
$\omega^{-2}$ \cite{KY,AK}.

Below we will show that the relationship (\ref{SKV}) is {\it
principally wrong}.
This means that one cannot obtain a correct
Fock contribution to the equilibrium persistent current by substituting
$\Re\sigma^{(2)}_{eq}({\bf q}; \omega)$ for the free-electron
nonlinear conductance $\sigma^{(2)}_{kin}({\bf
q},\omega,-\omega)$
(at
a finite momentum ${\bf q}$)
into Eq.(\ref{rf}).

To this end we use the expression \cite{KV} for
$\sigma^{(2)}_{eq}({\bf q}, i\omega_n)$ as the
derivative of the free-electron polarization operator $\Pi_0({\bf
q},i\omega_n)$
over the magnetic flux $\phi$:
\begin{eqnarray}\label{SF}
\sigma^{(2)}({\bf q}, i\omega_n) = \frac{c e^2}{2q^2}
\frac{\partial}{\partial \phi}
\overline{\Pi_0({\bf q},i\omega_n)},
\end{eqnarray}
where the overline denotes the disorder average and
\begin{eqnarray}\label{Pi}
\Pi_0({\bf q}, i\omega_n) =    T\sum_{{\bf k,k'},l}
G_{{\bf k}+{\bf q},{\bf k'}+{\bf q}}(i\epsilon_l + i\omega_n)
G_{{\bf k'},{\bf k}}(i\epsilon_l)\equiv
T\sum_{l}Tr_{{\bf k}}G(i\epsilon_l +
i\omega_n)\,G(i\epsilon_l).
\end{eqnarray}
Here $G$ are Matsubara Green's functions of non-interacting
electrons
in the presence of disorder; $\epsilon_l = (2l+1)\pi T$, and
the operation $Tr_{{\bf k}}$ is defined to shorten the following
expressions.

The analytical continuation in Eq.(\ref{Pi}) can be performed in a
general form:
\begin{eqnarray}\label{PiReal}
\Pi_0({\bf q},\omega)& =& Tr_{{\bf
k}}\int^{\infty}_{-\infty}
\frac{d\epsilon}{4\pi
i}\,
\left[[G^R(\epsilon + \omega)G^R(\epsilon) -
G^A(\epsilon)G^A(\epsilon -
\omega)]\tanh\left(\frac{\epsilon}{2T}\right)\right.+ \\
\nonumber & +&
2G^R(\epsilon + \omega)G^A(\epsilon)\left[n_F(\epsilon) -
n_F(\epsilon + \omega)]\right],
\end{eqnarray}
where $n_F(\epsilon) = [\exp{(\epsilon/T)} + 1]^{-1}$ and the retarded
($R$) and advanced ($A$) electron Green's functions are the analytical
continuations of
the Matsubara Green's function $G(\epsilon)$ into the upper
($\mbox{Im}\, \epsilon > 0$) and lower ($\mbox{Im}\, \epsilon < 0$)
half-planes, respectively.

The flux derivative
corresponds to
breaking of one Green's function into two by the current vertex
$\hat{j}$:
\begin{equation}
\label{fd}
\frac{\partial}{\partial \phi}\,G^{R(A)}(\epsilon) \rightarrow
G^{R(A)}(\epsilon)\hat{j}G^{R(A)}(\epsilon).
\end{equation}
It is important that neither the analytical properties nor frequency is
changed in the Green's functions involved.

Thus, the function in the r.h.s. of Eq.(\ref{SKV}) takes the form:
\begin{eqnarray}\label{SKV2}
&&\Re \,\sigma^{(2)}_{eq}({\bf q},\omega) \propto
Tr_{{\bf k}}\int^{\infty}_{-\infty} \frac{d\epsilon}{4\pi i}\,
\left[
[G^{R}(\epsilon+\omega)\hat{j}G^{R}(\epsilon+\omega)G^{R}(\epsilon)-
G^{A}(\epsilon)\hat{j}G^{A}(\epsilon)G^{A}(\epsilon-\omega)\right.+\\
\nonumber &+&
G^R(\epsilon + \omega)G^R(\epsilon)\hat{j}G^R(\epsilon) -
G^A(\epsilon)G^A(\epsilon - \omega)\hat{j}G^A(\epsilon - \omega)]
\tanh{\left(\frac{\epsilon}{2T}\right)} + \\ \nonumber
&+& \left.
2[G^{R}(\epsilon+\omega)\hat{j}G^{R}(\epsilon+\omega)G^{A}(\epsilon)+
G^R(\epsilon + \omega)G^A(\epsilon)\hat{j}G^A(\epsilon)]\,
[n_F(\epsilon) - n_F(\epsilon + \omega)] \right]+(\omega\rightarrow
-\omega).
\end{eqnarray}
Eq.(\ref{SKV2}) can be also obtained by
the direct analytical continuation of the expression
$K(i\omega_{l},i\omega_{m})$ given by the triangle
diagram in Fig.1a.:
\begin{equation}
\label{K-gen}
K(i\omega_{l},i\omega_{m})=Tr_{{\bf k}}T\sum_{n} G(i\epsilon_{n})
 G(i\epsilon_{n}+i\omega_{l})\hat{j} G(i\epsilon_{n}-i\omega_{m}).
\end{equation}
\begin{figure}[tbp]
\begin{center}
\psfig{file=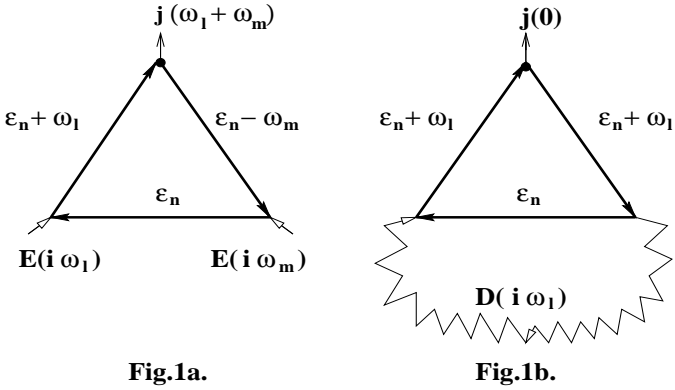,height=2.0in,width=3.5in}
\end{center}
\caption{a). A generic traingle diagram corresponding to Eq.(\ref{K-gen}).
Bold
lines denote electron Green's functions with Matsubare frequencies
indicated; b).The Fock contribution to an equilibrium persistent
current. The field vertices are connected with a photon Green's function. 
}
\end{figure}

The equilibrium Matsubara diagrammatic technique (see Fig.1b.) requires
$i\omega_{l}=-i\omega_{m}$, since the corresponding vertices
are connected by the photon Green's function $D({\bf q},i\omega_{l})$.
That is why in order to
obtain Eq.(\ref{SKV2}) one should {\it first set}
$i\omega_{l}=-i\omega_{m}$
in Eq.(\ref{K-gen}) and
{\it then} perform an analytical continuation of
$K(i\omega_{l})=K(i\omega_{l},-i\omega_{l})$
into
the
upper
half-plane of $\omega$. One can check that:
\begin{equation}
\label{ac-eq}
\Re\sigma^{(2)}_{eq}({\bf q},\omega) \propto
K^{R}(\omega)+K^{R}(-\omega)+c.c.
\end{equation}

The nonlinear conductivity $\sigma^{(2)}(\omega,-\omega)$ can also be
obtained from a generic triangle diagram Fig.1a. However, the proper
procedure is completely different.
It is determined by the {\it causal} nature of the system response to an
external field. For a generic quadratic response to a
time-dependent external field $E(t)$:
\begin{equation}
\label{resp}
I(t)=\int_{-\infty}^{\infty}d\tau_{1}\int_{-\infty}^{\infty}d\tau_{2}\,
\sigma(\tau_{1},\tau_{2})\,E(t-\tau_{1})\,E(t-\tau_{2}).
\end{equation}
causality requires that
$\sigma(\tau_{1},\tau_{2})\propto
\theta(\tau_{1})\theta(\tau_{2})$, where $\theta(x)$ is a step function.
This means that the generic nonlinear conductivity
$\sigma(\omega_{1},\omega_{2})$ is an analytical function
in the upper half-plane of {\it both} $\omega_{1}$ and $\omega_{2}$.

This is the key point in obtaining the nonlinear response from the
Matsubara diagrammatic technique \cite{AGD}. One should first find an
analytical continuation  $K^{(R,R)}(\omega_{1},\omega_{2})$ of a {\it
generic} triangle diagram
$K(i\omega_{l},i\omega_{m})$ into the upper half-plane of {\it both}
frequencies, and then set $\omega_{1}=-\omega_{2}=\omega$ to obtain a DC
nonlinear response:
\begin{equation}
\label{nlr}
\sigma_{kin}^{(2)}({\bf
q},\omega,-\omega)=
K^{R,R}(\omega,-\omega)+K^{R,R}(-\omega,\omega).
\end{equation}
By performing the analytical continuation in Eq.(\ref{K-gen}) we arrive at:
\begin{eqnarray}\label{S2Re}
&&\sigma^{(2)}_{kin}(\omega, -\omega) \propto
Tr_{{\bf k}}\int^{\infty}_{-\infty} \frac{d\epsilon}{4\pi i}\,  \left[
[G^R(\epsilon - \omega)G^R(\epsilon)\hat{j}G^R(\epsilon) -
G^A(\epsilon - \omega)G^A(\epsilon)\hat{j}G^A(\epsilon)]
\tanh{\left(\frac{\epsilon}{2T}\right)} + \right. \nonumber \\
& & \left. 2[G^A(\epsilon)G^R(\epsilon + \omega)\hat{j}G^A(\epsilon +
\omega)
- G^R(\epsilon)G^R(\epsilon + \omega)\hat{j}G^A(\epsilon + \omega)]
[n_F(\epsilon) - n_F(\epsilon + \omega)]
\right]+(\omega\rightarrow-\omega).
\end{eqnarray}
Now it is explicitly seen that the $R-A$ structure of
Eqs.(\ref{SKV2}),(\ref{S2Re})
is different even before disorder-averaging is done.
The quantity $\sigma^{(2)}_{eq}({\bf q},\omega)$ in
Eq.(\ref{SKV2}) has a structure of a derivative of the linear AC
polarizability with respect to the DC flux. Therefore the
there is no change of analytical properties in passing the current vertex.
On the contrary, the
nonlinear
conductance
$\sigma^{(2)}_{kin}(\omega, -\omega)$ is a derivative of the AC
conductance with respect to the AC flux.
Correspondingly, $R$ is changed to $A$ in passing the current vertex.
This proves invalidity of the basic assumption (\ref{SKV}) of Kopietz and
V\"olker Ref.\cite{KV}. The quantities Eqs.(\ref{SKV2}),(\ref{S2Re}) have
different physical meaning and the difference in their analytical
structure is crucial for the functional dependence of their
disorder-average values at high frequencies.

To this end, we note that only
such terms in Eqs.(\ref{SKV2}),(\ref{S2Re}) that contain both retarded
(R) and advanced (A) electron Green's
functions make a significant contribution to the disorder-averaged values.
This is because of the diffusion propagators (diffusons and
cooperons) $P({\bf q},\omega)\propto
Tr_{{\bf q}}\overline
{G^{R}(\epsilon+\omega)G^{A}(\epsilon)}$ which are
singular at
small $q$ and $\omega$. It implies that only last lines in
Eqs.(\ref{SKV2}),(\ref{S2Re}) are relevant.
Because $R$ is changed
to $A$ and frequency is unchanged in passing the DC current vertex in
Eq.(\ref{S2Re}),
it is possible  to
build a diffusion propagator (cooperon) of zero frequency in the kinetic
case. There is no such a possibility in the equilibrium
case, since the R-A structure is unchanged in passing the DC current
vertex.
The relevant two-cooperon diagrams are shown in Fig.2.
\begin{figure}[tbp]
\begin{center}
\psfig{file=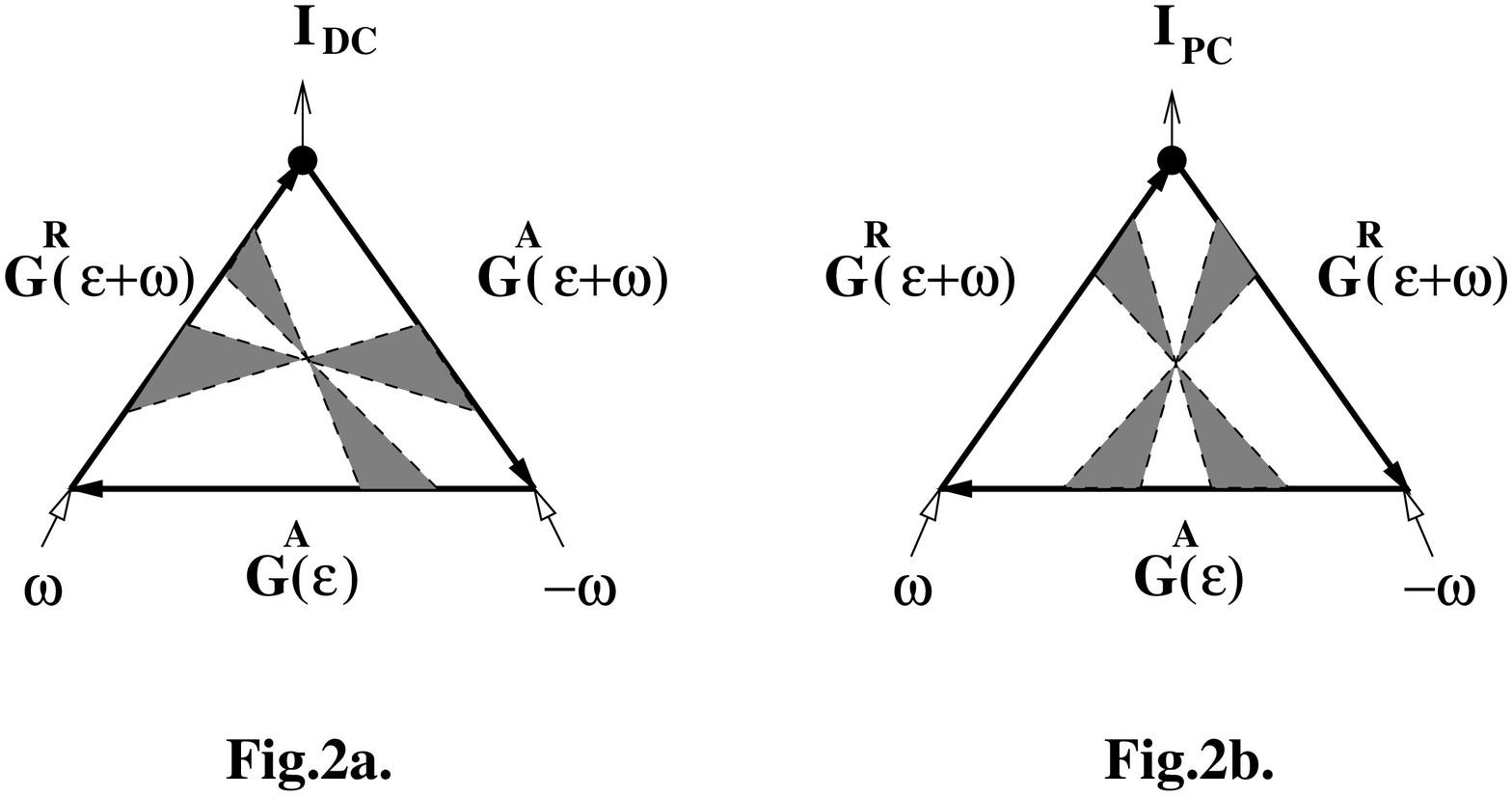,height=2.0in,width=3.5in}
\end{center}
\caption{a). The leading contribution to the disorder-average kinetic
coefficient $\sigma^{(2)}_{kin}(\omega,-\omega)$; b). The leading
contribution to the disorder-average value of
$\sigma_{eq}^{(2)}(0,\omega)$. 
The bold lines denote electron Green's functions; the shadowed
area between two dotted lines
denotes the fan series (cooperon) $P({\bf
q},\omega)$ with the
frequency $\omega$ equal to the difference of frequencies of electron
Green's
functions which are connected by the dotted lines.} 
\end{figure}
 
It turns out that both terms in the last lines of
Eqs.(\ref{SKV2}),(\ref{S2Re}) lead to identical contributions
to $\sigma^{(2)}_{kin}(\omega,-\omega)$ and
$\Re\sigma_{eq}^{(2)}(0,\omega)$, respectively.
In
the
case of nonlinear response where one of
the cooperons is of zero frequency,
the
summation over the cooperon momenta is mainly contributed by small
momenta $|k|\sim 1/L \ll 1/L_{\omega}=\sqrt{\omega/D}$. Then one can set
$k=0$ in the second cooperon and obtain a power-law
$\omega$-dependence. In the equilibrium case both cooperons are of the
same frequency $\omega$, since they appear due to differentiation
with respect to the DC flux of a single-cooperon diagram for the
Aharonov-Bohm effect in the linear
AC polarizability.
As a result, the diagram turns out to be
exponentially small in full agreement with the exponentially small
Aharonov-Bohm oscillations of the polarizability at high frequencies.

In conclusion, we have shown that the analytical structure of the
expression for the {\it free-electron nonlinear} conductance
$\sigma_{kin}^{(2)}({\bf q},\omega,-\omega)$ is different from that of the
coefficient $\sigma_{eq}^{(2)}({\bf q};\omega)$
that couples the Fock contribution to the equilibrium persistent current
and the spectral density of the electric field fluctuations in
the system of {\it interacting}
electrons.
The difference is caused by the {\it causality} requirement
that determines the nonlinear response to an external field
but is irrelevant for the interaction contribution to
equilibrium quantities.
As a result, at high frequencies
$\omega\gg E_{c}$ the free-electron nonlinear conductance
$\sigma_{kin}^{(2)}(\omega,-\omega)\propto 1/\omega^2$, while
$\sigma_{eq}^{(2)}({\bf q}\rightarrow 0;\omega)$ is exponentially
small.
This analysis  warns against a possible
mistake when one applies methods developed for
describing kinetic phenomena to equilibrium systems and
{\it vice versa}.

\end{document}